\documentstyle[prl,multicol,aps,graphics]{revtex}
\begin{document}
\draft
\title{Linear scaling relaxation of the atomic positions in nano structures}
\author{Stefan Goedecker, Fr\'ed\'eric Lan\c{c}on, Thierry Deutsch }
\address{D\'epartement de recherche fondamentale sur la mati\`ere condens\'ee,\\
         SP2M/NM, CEA-Grenoble, 38054 Grenoble cedex~9, France}
\date{\today}
\maketitle
\begin{abstract}
We present a method to determine the equilibrium geometry of large 
atomistic systems with linear scaling. It is based on a separate 
treatment of long and short wavelength components of the forces.
While the rapidly varying part is handled by conventional methods, 
the treatment of the slowly varying part is based on elasticity 
theory. As illustrated by numerical examples containing up to 
a million atoms this method allows 
an efficient relaxation of large nanostructures.

\end{abstract}

\begin{multicols}{2}
\setcounter{collectmore}{5}
\raggedcolumns

Considerable effort has recently been devoted to the development of 
various linear scaling algorithms. Only with these types of algorithms it 
is possible to perform atomistic simulations of large systems containing 
many atoms. Due to these efforts it is nowadays possible to calculate 
the total energy of the system and the forces acting on the atoms with 
linear scaling under many circumstances. Linear scaling can obviously been 
obtained if short range 
force fields are used. With the help of sophisticated algorithms such as 
the fast multipole methods~\cite{fmp} and particle mesh methods it is also possible 
to obtain linear scaling for force fields that include long range electrostatic 
interactions. Finally linear scaling can be obtained if the forces are 
calculated quantum mechanically by using so called O(N) electronic structure 
methods~\cite{on}. 

In spite of this important algorithmic progress concerning length 
scaling problems various time scaling problems persist. 
Ordinary molecular dynamics 
simulations can only cover relatively short time interval's. Recently 
developed schemes such as hyper-dynamics~\cite{voter} can detect rare event 
on much longer time scales than previously possible. Another related 
scaling problem is encountered in geometry optimizations of large 
atomistic systems. Geometry optimizations require the minimization 
of the total energy with respect to the atomic positions. 
The number of iterations required by standard minimization methods 
such as the conjugate gradient method increases with system size, 
destroying thus linear scaling even if the forces are calculated with 
such a scaling.

Even though the problem of the increasing number of iterations in 
geometry optimizations has been observed by many workers in the field, 
it has to the best of our knowledge not been 
analyzed up to now. We will therefore start with an detailed 
examination of the effect and relate it to well known facts about 
the convergence rate of iterative methods before presenting our 
solution how to overcome it.

\begin{figure}[ht]
     \begin{center}
      \setlength{\unitlength}{1cm}
       \begin{picture}( 5.,2.2)           
        \put(-1.5,0.0){\includegraphics{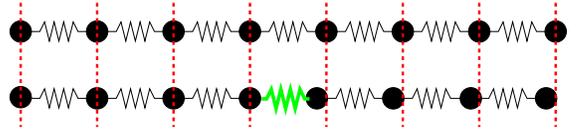}}   
       \end{picture}
 \caption[]{\label{chain}
	   A linear chain in equilibrium (top) and in a special 
           out-of-equilibrium configuration (bottom).} 
      \end{center}
     \end{figure}

Figure~\ref{chain} shows a linear chain, where  the atoms are connected 
by elastic springs. The upper panel shows the equilibrium configuration, 
the lower one a configuration where the right half is shifted to the 
left, compressing the spring in the middle. Let us now consider 
what happens if we use the lower configuration as the starting point for 
a geometry optimization using any standard method such as the conjugate 
gradient or a quasi-newton method. Because only the forces acting 
on the two atoms neighboring the compressed spring are non-vanishing, 
we will relax only these two atoms in the middle in the first iteration. 
In the subsequent iterations non-vanishing forces will appear on all 
the atoms whose neighbors have been moved in previous iterations and 
hence they will be moved as well. Consequently it takes $n/2$ iterations 
for a chain of length $n$ to propagate the perturbation from the 
compressed spring in the middle to the end. It follows that one needs 
at least of the order of $n$ iterations to find the equilibrium configuration.

One can recover the basic result of this intuitive derivation from 
a mathematical analysis. Writing down the elastic energy of a periodic 
three-dimensional model 
solid connected by perfect springs one finds that the eigenvalues $\lambda_i$ 
of the Hessian matrix are the squares of the phonon frequencies $\omega_i$.
Mathematical theorems about the convergence of iterative methods ~\cite{saad} 
relate the number of iterations to the condition number $\kappa$, which is 
the ratio between the largest and the smallest eigenvalue of the Hessian 
matrix. Considering the case of large condition numbers $\kappa$ and 
assuming that in the displacements that would bring the initial 
non-equilibrium configuration back into the equilibrium configuration all 
wavelengths are roughly equally represented, the number of
iterations $n_{it}$ for the conjugate gradient method is given by
\begin{equation}
n_{it}^{CG} \propto \sqrt{\kappa} | \log(\epsilon) |
           = \sqrt{\frac{\lambda_{max}}{\lambda_{min}}} | \log(\epsilon) | 
           = \frac{\omega_{max}}{\omega_{min}} | \log(\epsilon) |  \label{nit}
\end{equation}
where $\omega_{min}$ is the smallest non-zero phonon frequency and 
$\epsilon$ the required precision. The highest phonon frequency 
$\omega_{max}$ arising either from an optic or acoustic branch is nearly 
independent of system size, whereas the acoustic phonon branch goes 
to zero at ${\bf k}= 0$ linearly. The grid of allowed wave vectors ${\bf k}$
in the Brioullin zone becomes 
finer and finer as the periodic cell of the solid grows. 
The smallest phonon frequency $\omega_{min}$ is related to the grid point 
that is closest to the origin and is proportional to $1/L_{max}$ 
where $L_{max}$ is the largest linear dimension of the system. The number 
of iterations is therefore given by

$$ n_{it}^{CG} \propto \frac{L_{max}}{a} | \log(\epsilon) |   $$

where $a$ is a typical inter-atomic distance. 

For the steepest descent method the convergence is slower:  
\begin{equation}
n_{it}^{SD} \propto \frac{\lambda_{max}}{\lambda_{min}} | \log(\epsilon) |
           = \frac{\omega_{max}^2}{\omega_{min}^2} | \log(\epsilon) | \nonumber
\end{equation}

Even though these results were derived for periodic solids numerical 
simulations clearly show that they also hold true for regular non-periodic 
structures such as linear polymers, two-dimensional sheets and 
bulk like clusters. 

It is interesting to relate the number of conjugate gradient 
iterations to the number of atoms $N$ 
in the system. For a one-dimensional system we find (Eq.~\ref{nit}) 
that $n_{it}^{CG}$ is 
proportional to the number of atoms $N$, in the two-dimensional case 
$n_{it}^{CG} \propto N^{1/2}$ and in the three-dimensional case 
$n_{it}^{CG} \propto N^{1/3}$ assuming that all the side lengths are comparable.
The possible gains by using a linear scaling algorithm are thus largest 
in the one-dimensional case.

In chemistry, geometry optimizations are usually done in 
internal coordinates (see~\cite{nemeth,pulay} and references therein) 
instead of Cartesian coordinates. An extension of this approach to 
periodic systems has also been investigated~\cite{scuseria}. The 
internal coordinates typically consist of bond stretching, bending 
and torsion coordinates. Their construction is not easy for complicated 
geometries and  transformations between the Cartesian and 
internal coordinates are required in each iteration step. 
These transformation are costly and have a cubic scaling when done 
by conventional linear algebra methods. Recently ways have however 
been found to do these transformations with linear scaling~\cite{nemeth}.
For the example in Fig~\ref{chain} there would be no dependence 
of the number of iterations on the size of the system if the 
optimization was done in internal coordinates (just bond lengths in 
this case). This property is also 
satisfied for simple polymers, but is lost when one has molecules 
with a higher dimensional character or bulk like structures.

\begin{figure}[ht]
     \begin{center}
      \setlength{\unitlength}{1cm}
       \begin{picture}( 5.,3.0)           
        \put(-0.0,.0){\includegraphics{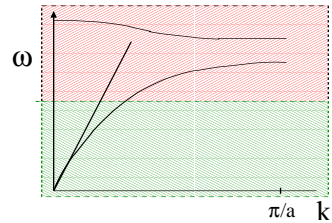}}   
       \end{picture}
 \caption[]{\label{divcon}
            Illustration of the subdivision into a high frequency 
            part treated by conventional methods and a high frequency 
            part described by elasticity theory. The  straight line 
            is the "phonon" dispersion relation of an elastic solid 
            in the continuum limit.}
     \end{center}
      \end{figure}

Linear scaling is generally achieved by treating different length 
and time scales in an approriate ways. This will also be the guiding 
principle in this work. As we have seen in the preceeding discussion the 
slowdown of the convergence rate is due to the long wavelength 
acoustic phonons. Hence the basic idea is to treat 
high and low frequencies in a different way. The reduction of 
the high frequency components of the force during an geometry 
optimization is satisfactory with the 
standard methods such as conjugate gradient and steepest descent.
To reduce the low frequency components we will 
use elasticity theory. This is motivated by the fact that ordinary 
materials have continuum like behavior on length scales of just 
a few inter-atomic distances. This divide and conquer approach in 
frequency is illustrated in Figure~\ref{divcon}.

Elasticity theory, discretized on a grid of $M$ grid points, 
gives the total elastic energy $E_{el}$ of a system as a 
quadratic form of the $3 M$ displacements $u_i$ 
from the equilibrium configuration. 

$$ E_{el} =  \frac{1}{2} {\bf u}^T {\bf A} {\bf u} $$

Differentiating  with respect to ${\bf u}$ and using the fact 
that the force ${\bf f}$ is just the negative of the 
derivative of the energy with respect to 
the atomic positions we obtain 

\begin{equation} 
 - {\bf A} {\bf u} = {\bf f}  \label{auf} 
\end{equation}

Eq.~\ref{auf} is the central equation for the relaxation steps of the 
minimization iteration that are based on elasticity theory. Given 
the forces it allows us to calculate the displacements $-u_i$ that would 
bring the solid into its equilibrium configuration. An ideal harmonic 
solid can be relaxed with one single step in this way. If we apply 
this step as explained below to our real solid it drastically 
brings down the low frequency components of the force. 

Eq.~\ref{auf} cannot be applied in a straightforward way to an atomic system. 
We have only the forces acting on the atoms but not forces acting 
on grid points. In the same way we would like the displacements for 
the atoms and not for the grid points. 
Methods to transform between atoms and grid points are well known 
form the various particle mesh methods. Following these ideas we 
map the atomic forces onto the grid by smearing them 
out with the function $(1-x^2/8)^4 (1-y^2/8)^4 (1-z^2/8)^4$ 
(the unit length is the grid spacing).
The transformation from the grid onto the atoms is done by 
cubic interpolation. 

Just by invoking elasticity theory we have actually not yet really solved the 
scaling problem. If we were to solve Eq.~\ref{auf} by a conjugate gradient 
method the number of iterations would as well increase with respect to the 
number of grid points. In this case the method might still be useful 
since under most circumstances a conjugate gradient step for Eq.~\ref{auf} 
will be significantly less expensive than for the atomic system, but the 
overall scaling would not be linear. Fortunately recent developments 
in the field of multi-grid methods~\cite{brezina} allow us to solve 
sparse linear systems of the type of Eq.~\ref{auf} with linear scaling 
and with small prefactors. In the case 
of periodic systems one can also use FFT techniques that exhibit a 
nearly  linear $M \log(M)$ scaling. In summary, if we can calculate the 
forces with linear scaling and solve Eq.~\ref{auf} with linear scaling 
using the above mentioned methods overall linear scaling can be obtained 
for the geometry optimization problem in Cartesian coordinates. 

The complete algorithm consists of a iteration of the following steps. 

\begin{itemize}
\item Perform $n_{c}$ conventional minimization sweeps.
      The choice of $n_{c}$ is discussed below.
\item Perform one step where the solid is described by 
      elasticity theory.
\begin{itemize}
\item Calculate the forces acting on the atoms
\item Transfer the atomic forces onto a computational grid
\item Solve Eq.~\ref{auf} with a linear scaling method to obtain 
      the displacements
\item Evaluate the numerical displacement field at the atomic positions 
      to get the atomic displacements.
\item  Move the atoms along these displacement directions. This can either 
      be done using a fixed step size or by a line minimization along 
      these directions. We choose the later variant.
\end{itemize}
\end{itemize}

\begin{figure}[ht]
     \begin{center}
      \setlength{\unitlength}{1cm}
       \begin{picture}( 5.,4.1)           
        \put(-1.5,-.5){\includegraphics{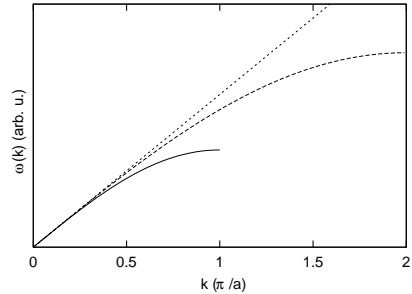}}   
       \end{picture}
 \caption[]{\label{disrel}
         The "phonon" dispersion relation of two grids used 
         for the solution of Eq.~\ref{auf}. The wave-vector $k$ 
         is measured in units of $\pi/a$, where $a$ is the lattice constant 
         of the physical system. The solid line corresponds to 
         a grid whose spacing is equal to the inter-atomic distance $a$, 
         the dashed line to the case where the grid-spacing is half of 
         the inter-atomic distance, and the dotted line is the continuum limit.} 
     \end{center}
\end{figure}

Continuum elasticity theory considers the limit where the ratio 
between the wavelength of any perturbation is much larger than the 
inter-atomic spacing. In this limit the phonon dispersion relation 
is a straight line as shown in Figure~\ref{disrel}. 
The point at which the phonon dispersion relation of the true solid 
starts to deviate from a straight line tells us the length scale 
at which the solid can already be considered as a continuous elastic 
medium. Once one has 
a discretized version of elasticity the "phonon" dispersion 
relations are not any more straight lines but the phonon dispersion 
relation of a model harmonic solid. The size of their Brioullin 
zone depends on the density of the numerical grid as shown in 
Fig.~\ref{divcon} and~\ref{disrel}.

The number of conventional sweeps $n_{c}$ in our method 
depends on the size of the frequency range that the conventional part 
needs to cover. This range depends both on the 
physical properties of the system and on the density of the numerical 
elasticity grid. Obviously the range will be larger if there are high 
optical phonon bands. It will also be significantly larger if the number of 
grid points is less than the number of atoms since in this 
case the highest frequency that can be reached by the elastic model 
is small. According to our experience 
it is best to choose the number of grid points $M$ to be in the interval 
[$N$ : $2^d$ $N$], where $d$ is the dimension. 
Increasing $M$ within this range reduces the number of 
iterations but each iteration becomes more expensive since the solution 
of Eq.~\ref{auf} takes longer. Increasing $M$ further does not 
lead to a significant additional reduction of the number of iterations.
Choosing $M$ smaller than $N$ is not reasonable since for such 
values the number of iteration increases strongly and
the effort for the solution of Eq.~\ref{auf} is completely negligible.
We found that for grid densities in the recommended interval 
the best values for $n_{c}$ were between 2 and 4. For silicon  
systems conjugate gradient sweeps were slightly more efficient than 
steepest descent sweeps.

Even though each of the two steps in our method targets a specific frequency 
range a clearcut separation is not possible. The conjugate 
gradient step will slightly dampen the low frequency component.
The elasticity step will introduce some additional short wavelength 
errors since the "phonon" dispersion relation for the elastic grid model 
deviates from the phonon dispersion relation of the material in the 
high frequency region. These errors are however very small since these 
components were already significantly decimated by the preceding 
conventional steps and they are therefore immediately 
annihilated by the following conjugate gradient steps. 

Table~\ref{divac} shows numerical results for the relaxation of 
a divacancy in silicon crystals of varying cell size modeled  with the 
EDIP inter-atomic potential~\cite{kaxiras} The norm of the forces 
$\sqrt{\sum_i f_i}$ was reduced down to 10$^{-8}$ resulting in 
total energies that were converged to machine precision. The 
initial configuration was the perfect silicon crystal with 
two neighboring atoms removed. We compare 
both the number of force evaluations and the CPU time of the 
conjugate gradient method with our method. All line minimizations were 
done by minimizing the projected force and under the assumption that 
the energy function is quadratic. This results in just two force evaluations 
per conjugate gradient step and turned out to be the most efficient 
conjugate gradient implementation.  The elastic equations of 
an isotropic homogeneous medium were solved using Fourier techniques.

Taking the CPU time as the criterion, the crossover-point is at 
around 1000 atoms for EDIP force field that is fast to evaluate. 
If we would use another total energy scheme 
where the calculation of the forces is more costly, 
the number of force evaluations would be the best criterion and the 
crossover-point goes down to 600 atoms roughly. 

The divacany example allows the examination of the scaling behavior 
over a large range of system sizes, since the behavior of the 
divacancy remains qualitatively the same. In many other nanosystems 
the behavior can fundamentally change with size due, for instance, to size 
dependent surface reconstructions. This can then lead to highly 
irregular trends in the number of iterations. 
For these reasons we have choosen the divacany example even though it is a 
particularly difficult example to demonstrate the advantages of the 
linear scaling method over the conjugate gradient method. It is fully 
three-dimensional and the importance of the long wavelength perturbations 
induced by the di-vacancy decreases with system size, leading to an increase 
of the number of conjugate gradient iterations that is slighly slower 
than predicted by Eq.~\ref{nit}. 

\begin{table}[b]
\caption[]{ Number of force evaluations $n_f$ and CPU time $T$ in seconds 
for the conjugate gradient (CG) and the linear scaling (SC) 
method for a divacancy in silicon. \label{divac}}
\begin{tabular}{|l||c|c|c|c|} \hline
 number of atoms  & $n_f$(CG) & $n_f$(LS) & T(CG) & T(LS) \\
\hline 
  510      & 102  & 106  & .41  & .50  \\ \hline
  998      & 124  & 106  & .90  & .93  \\ \hline
  1726     & 146  & 109  & 1.7  & 1.6  \\ \hline
  4094     & 184  & 115  & 5.1  & 4.2  \\ \hline
  13822    & 260  & 115  & 24.  &  14. \\ \hline
  110592   & 502  & 115  & 373.  & 135.  \\ \hline
  884734   & 934  & 117  & 5586. & 1147. \\ 
\end{tabular}
\end{table}

We have also applied the method to 
more complicated systems such as incommensurate interfaces in silicon 
and clusters of silicon. We could also in these application find a significant 
reduction in the number of necessary force evaluations compared to the 
conjugate gradient method. As expected the gains were particularly large for 
systems with two-dimensional character. Our previous analysis of the number of 
iterations for geometry optimization was based on the assumption that we 
are in a region were the energy functional has nearly quadratic behaviour. 
Our numerical experiments indicate however that our linear scaling method 
also offers advantages in the case where one starts the geometry optimization 
from a point that is far away from such a region. In this case multiple minima 
are frequently encountered. We found that our linear scaling method is more 
likely to find lower energy local minima than the conjugate gradient method.

We have presented a method that allows to relax very large systems 
with linear scaling, removing thus an important bottleneck in 
atomistic simulations. This method in combination with force 
field and O(N) electronic structure methods makes for the first time possible 
efficient geometry optimizations for large solid state 
materials and nanostructures. It is expected that it can 
be extended to large molecular and biological systems. 

S.G. thanks M. Brezina, K. Nemeth P. Pulay and G. Scuseria 
for interesting discussions

\end{multicols}{2}

    \end{document}